\documentclass{article}

\usepackage{arxiv}

\usepackage[utf8]{inputenc} % allow utf-8 input
\usepackage[T1]{fontenc}    % use 8-bit T1 fonts
\usepackage{hyperref}       % hyperlinks
\usepackage{url}            % simple URL typesetting
\usepackage{booktabs}       % professional-quality tables
\usepackage{amsfonts}       % blackboard math symbols
\usepackage{nicefrac}       % compact symbols for 1/2, etc.
\usepackage{microtype}      % microtypography
\usepackage{lipsum}
\usepackage{times}
\usepackage{epsfig}
\usepackage{amssymb}
\usepackage{relsize}
\usepackage{enumerate}	%For enumeration
\usepackage{epstopdf}	%Converts eps files on the fly to pdf for inclusion in pdf
\usepackage{array}		%For more advanced tables
\usepackage{import}		%Allows using subimport instead of include (see details below)
\usepackage{bytefield}
\usepackage{fixme}
\usepackage{makecell}
\usepackage{tabulary}
\usepackage{xcolor}
\usepackage{graphicx}
\usepackage{algorithmic}
\usepackage{bm}
\usepackage{float}
\usepackage{hhline}
\usepackage{pdfpages} 
\usepackage{siunitx}
\usepackage{enumitem}
\usepackage{caption}
\usepackage{amsmath}

\DeclareMathOperator*{\argmax}{argmax}

\newcommand{\degreee}{^{\circ}}
\newcommand{\omni}{$360\degreee${ }}

\title{HSMF-Net: Semantic Viewport Prediction for \\ Immersive Telepresence and On-Demand 360-degree Video}

\author{
  Tamay Aykut \\
  Stanford University\\
  \texttt{tamaykut@stanford.edu} \\
   \And
 Basak Gülecyüz \\
  Technical University of Munich\\
  \texttt{basak.guelecyuez@tum.de} \\
  \And
 Bernd Girod \\
  Stanford University\\
  \texttt{bgirod@stanford.edu} \\
  \And
 Eckehard Steinbach \\
  Technical University of Munich\\
  \texttt{eckehard.steinbach@tum.de} \\
}

\begin{document}
\maketitle

\begin{abstract}
The acceptance of immersive telepresence systems is impeded by the latency that is present when mediating the realistic feeling of presence in a remote environment to a local human user. A disagreement between the user's ego-motion and the visual response provokes the emergence of motion sickness. Viewport or head motion (HM) prediction techniques play a key role in compensating the noticeable delay between the user and the remote site. We present a deep learning-based viewport prediction paradigm that fuses past HM trajectories with scene semantics in a late-fusion manner. Real HM profiles are used to evaluate the proposed approach. A mean compensation rate as high as 99.99\% is obtained, clearly outperforming the state-of-the-art. An on-demand \omni video streaming framework is presented to prove its general validity. The proposed approach increases the perceived video quality while requiring a significantly lower transmission rate.
\end{abstract}

% keywords can be removed

\section{Introduction}\label{sec:introduction}
\begin{figure}[t]
                \centering
                \begin{tabular}{c}
                                \includegraphics[width=.7\textwidth]{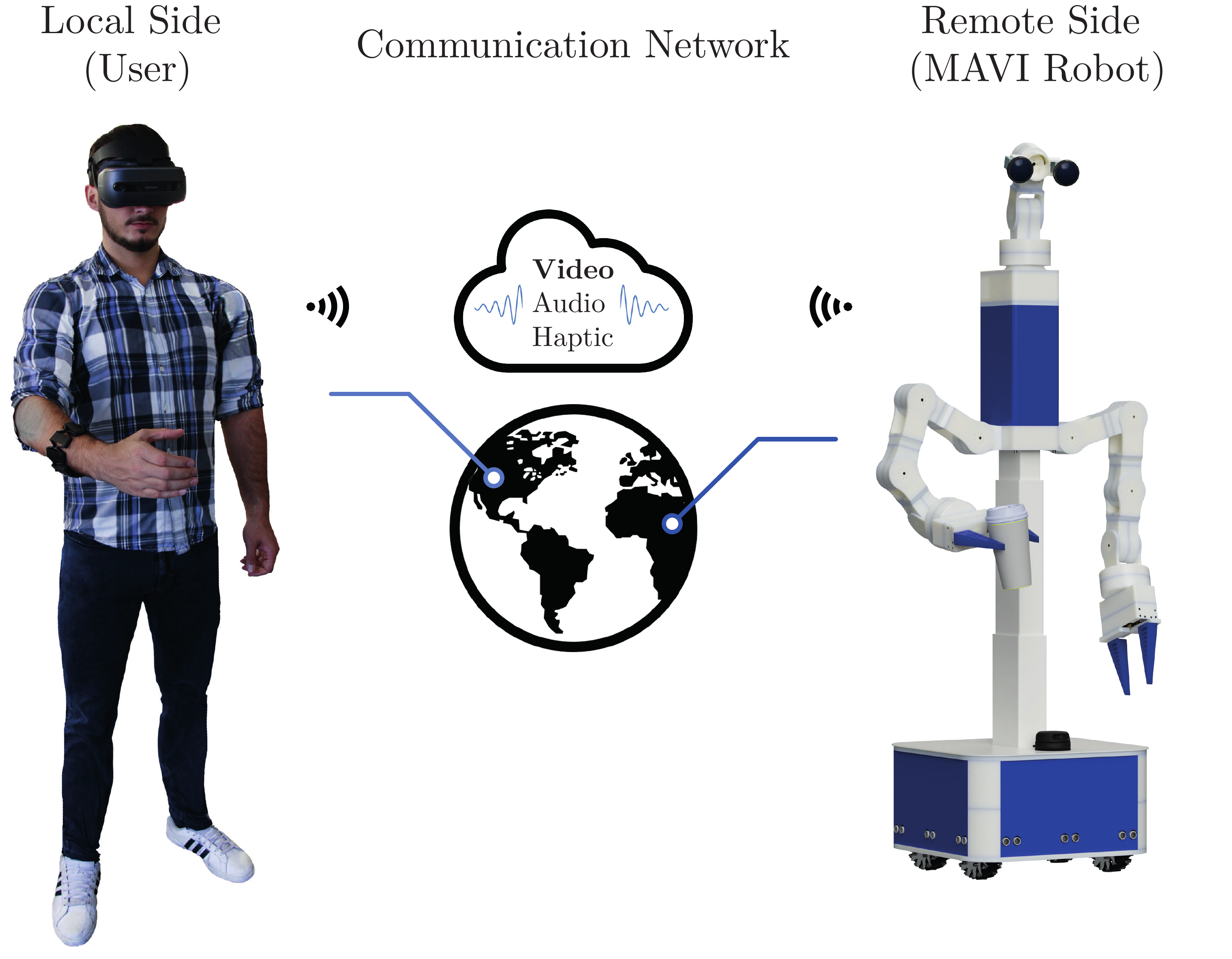}\\
                                %(a) \\ 
                                \includegraphics[width=.7\textwidth]{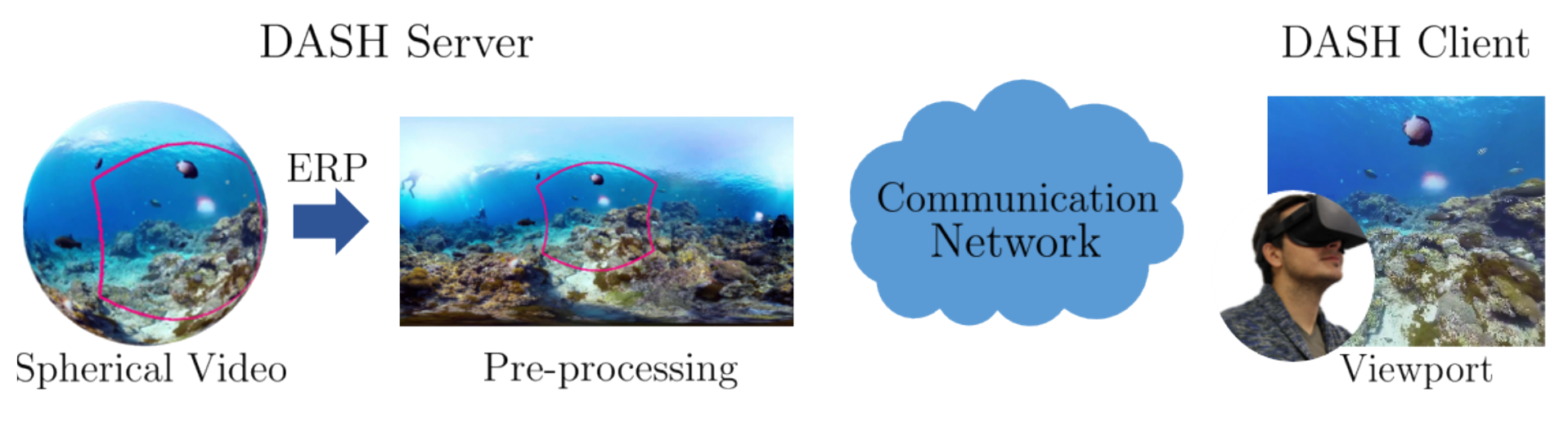}\\
                                %(b) \\ 
                \end{tabular}
                \caption{Top: Overview of a networked telepresence scenario. Bottom: Streaming pipeline for immersive \omni videos.}
                \label{fig:telepresence-streaming} 
\end{figure}

Immersive telepresence is the ability to virtually transport humans from one place to another instantly. Provisioning such technology in high-fidelity finds promising usage in forthcoming applications such as teledriving, telemaintenance, and remote healthcare. A server-/client-based architecture can be built to allow human users to immerse themselves into a distant environment upon request as shown in Fig.~\ref{fig:telepresence-streaming}. Head-mounted displays (HMDs) are widely deployed as a mean to increase the level of immersion. The mediation of omnidirectional visual data has a sizable impact on the perceived quality of experience (QoE)~\cite{Hen96}. Physically unavoidable delays cause noticeable lag between head-motion (HM) and visual response. The user’s perceived ego-motion needs to comply with the sensory impressions from the visual system, the vestibular system, and the non-vestibular proprioceptors \cite{Wat08}; a mismatch provokes the emergence of motion sickness.
Sending a 360$\degreee\times$180$\degreee$ video would allow the user to explore the scene not noticing the lag given the immediate access to the entire visual representation. This results, however, in large transmission rates, while a considerable portion of the imagery remains unused as only the user’s viewport is rendered onto the HMD. Sending a single segment of the 360$\degreee$ video and forecasting successive segments is more efficient in terms of smooth and non-disturbing view transitions.
Providing a stereoscopic \omni visual representation of the distant scene further fosters the level of realism and greatly improves task performance.
Omnistereoscopic vision allows the user to perceive depth information and sense the distant scene in 3D.
The realtime acquisition and streaming of \omni stereo vision faces major challenges. State-of-the-art (SotA) solutions are bulky, not realtime, and tend to produce erroneous output due to the stitching processes involved, which perform poorly for texture-less scenes. Aykut et al.~\cite{TA19} proposed an actuated 3D 360$\degreee$ vision system with delay compensation that performs live telepresence upon request. The buffer-based delay compensation approach significantly decreases the perceived motion-to-photon (M2P) latency.
Proper viewport (or HM) prediction is needed for immersive user experience both for live telepresence and on-demand 360$\degreee$ video streaming. Erroneous prediction can lead to wrong viewport selections and is detrimental to visual comfort.
This paper presents a viewport prediction paradigm that is useful for both immersive telepresence and on-demand 360$\degreee$ video streaming. The forecasting is not only based on past head trajectories but also leverages scene content for proper prediction, which is denoted as \textit{semantic viewport prediction}. The core idea is to provide highly precise prediction capabilities using HM data, which proved to perform fairly well, and augment the predictability with spatio-temporal scene information by means of saliency and motion maps.
A late-fusion policy is presented that is robust against poor lighting conditions and other  disturbances.  The contributions of this paper can be summarized as follows:
\begin{itemize}[leftmargin=1em]
\item A novel HM prediction scheme is proposed that extends previous techniques by spatio-temporal scene information. Scene semantics are merged with past HM data by means of a novel deep learning-based late fusion technique.
\item The novel prediction paradigm is examined for both realtime immersive telepresence, exhibiting a high mean delay compensation rate of 99.99\%, and on-demand 360$\degreee$ video streaming, greatly improving the perceived video quality while claiming significantly lower transmission rates.
\end{itemize}
%------------------------------------------------------------------------
\section{Related Work}
HM prediction is a valid method to improve the performance of streaming applications. Rather than sending the current HM value, the prospective HM is estimated with respect to the present end-to-end delay. Predicting and sending the future head position helps to decrease the M2P-latency.
 
One way to forecast HMs is to exclusively consider the past head trajectory and fit a first-order polynomial employing (linear) regression \cite{Mav10, Seb12}. 
Another way is to use (Extended) Kalman or Particle filters to obtain an optimal head state estimate in terms of head orientation, angular velocity, and acceleration. The estimated state is then employed into a motion model for linear or polynomial extrapolation. They are often sensitive to noise, suffer from erroneous prediction, and tend to overshoot for quick motions and abrupt orientation changes. Recent approaches employ supervised learning techniques to improve the fixation prediction \cite{Cha16}.
 
These methods solely rely on HM sensory data. An alternative approach is to leverage the image content to detect the user's region of interest. Saliency maps can be created to indicate the user's visual attraction and eye fixation likelihoods. The temporal scene component can be further incorporated to improve the overall performance. Xu et al. \cite{Xu18} combined a head trajectory encoder module and a saliency encoder module. Both modules are early fused with fully connected dense layers to predict the change in HM.
 
As opposed to the SotA, this paper emphasizes the HM data as the main source of information and exploits spatio-temporal scene semantics to further optimize the forecasting capabilities. The performance of the proposed approach is thereby agnostic to poor image data caused by bad lighting, noisy or distorted capture, etc . The presented deep learning-based late fusion model is based on stacked gated recurrent units and convolution layers to extract the most distinct features at different granularities. The proposed network is realtime-capable and outperforms the SotA.
%------------------------------------------------------------------------
\section{Proposed Approach}
The essential pillar of the proposed approach is a deep architecture based on past head trajectories (\textit{H-network}), which is enhanced with scene semantics. Two further networks are proposed and separately trained on spatial and temporal image information by means of saliency (\textit{S-network}) and motion maps (\textit{M-network}). A late fusion network (\textit{F-network}) is presented that merges the individual components to one prediction scheme that is agnostic to noisy, low-quality images, poor lighting conditions, and any other potential source of disturbance.
\begin{figure*}[h]
	\centering
	\begin{tabular}{ccc}
		\includegraphics[width=.25\textwidth]{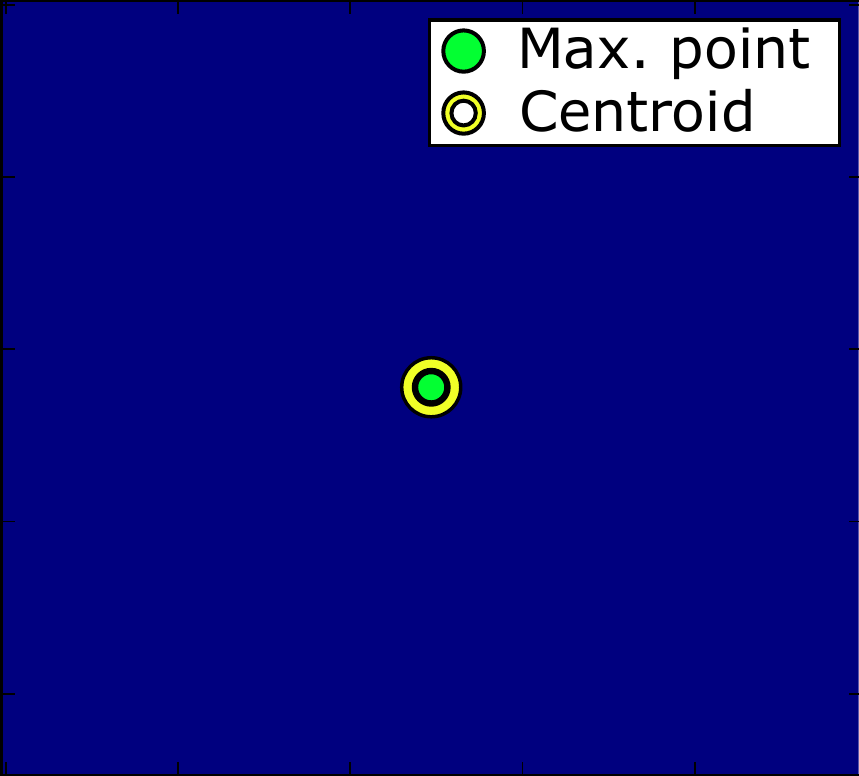} &
		\includegraphics[width=.25\textwidth]{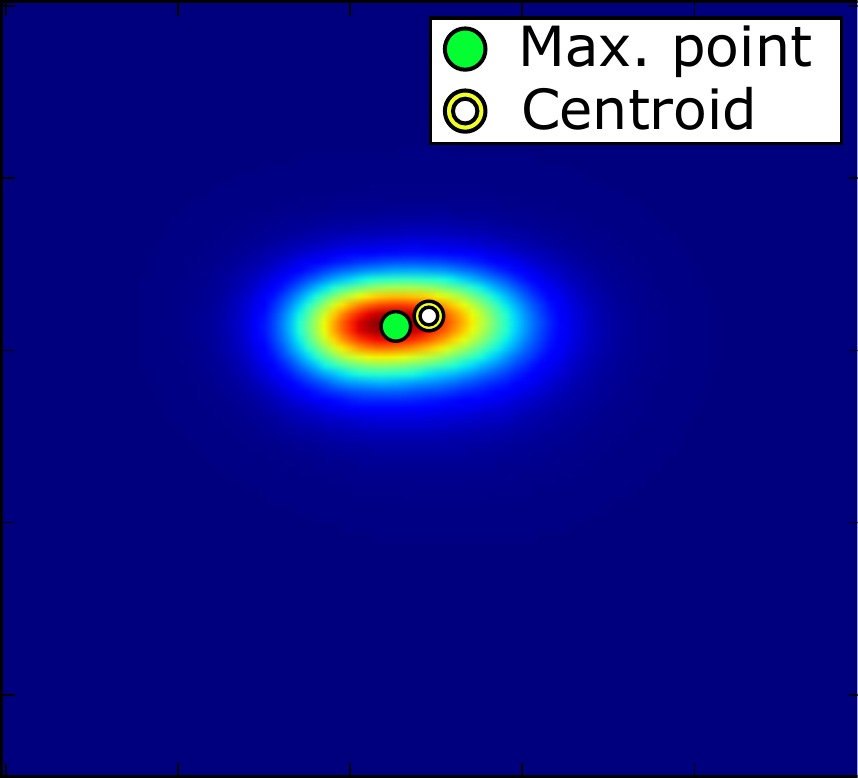} & 
		\includegraphics[width=.25\textwidth]{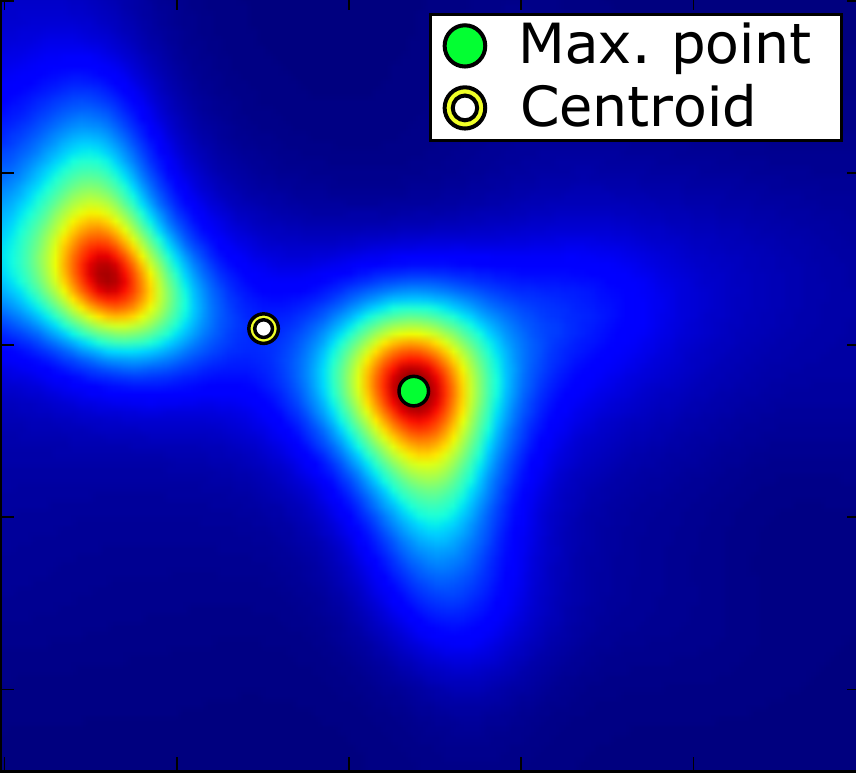}
	\end{tabular}
	\caption{The max- and centroid-based exploitation techniques are demonstrated for different saliency maps.}
	\label{fig:saliency-moments}
\end{figure*}
\subsection{Head Orientation-based Network (H-network)}
The deep network from \cite{TA19}, which outperformed the SotA, is adopted for HM-based viewport prediction.

The input of the network is the user's current head state $\bm{\xi}=[\theta\hspace{.5em}\phi\hspace{.5em}\psi]^{T}$ being the pan $\theta(t)$, tilt $\phi(t)$, and roll $\psi(t)$ orientations at time $t$ along with their past trajectories within a time window (W=250ms, $\Delta t = 12.5$ms). To ensure fully generalizable inference, the input is first scaled to $[-180\degreee;180\degreee]$ and converted into the differential domain as per:
\begin{equation}
\bm{\xi}_{\text{diff}}(t)=\bm{\xi}(t)-\bm{\xi}(t-\mathsmaller{\Delta}t).
\end{equation}

The differences are then normalized with: 
\begin{equation}
\tilde{\bm{\xi}}_{\text{diff}}(t) = \frac{\bm{\xi}_{\text{diff}}(t)}{\max(|\bm{\xi}_{\text{diff}}|)}
\end{equation}

to be in the range of $[-1;1]$. The output of the deep network predicts a whole course of future orientations from 0.1--1s. The actual output of the network, however, is a sequence of future normalized differences $\hat{\tilde{\bm{\xi}}}_{\text{diff}}(t)$. The remapping of normalized differences to absolute orientation values is accomplished with:
\begin{equation}\label{eq:remapping}
\hat{\bm{\xi}}(t) = \bm{\xi}(t) + \sum_{k = t - \tau/\mathsmaller{\Delta}t}^{t} \hat{\tilde{\bm{\xi}}}_{\text{diff}}(k) \cdot \max\left(\left|\bm{\xi}_{\text{diff}}\right|\right).
\end{equation}
\subsection{Saliency Network (S-network)}
Saliency maps $S(x, y, t)$ convey information about the visual attractiveness of spatial regions within an image frame as shown in Fig.~\ref{fig:saliency-moments}. They can contain information that might help to improve the prediction accuracy. A user experiencing a \omni video with an HMD is likely to navigate to regions that might appear visually attractive as head and eye gaze interactions are coupled and gaze statistics are correlated with saliencies within the scene \cite{Sit18}.
The Deepgaze II \cite{Kum17} model is adopted to create saliency maps given its high score rating, especially for the AUC metric, which interprets saliency maps as classifiers for pixels that are fixated or not \cite{Byl16}.
The S-network considers all saliency maps that are within an empirically determined window $W_c=\SI{500}{\milli\second}$. Given the sparse nature of saliency maps, we decided to manually extract scene semantics. The (1) max- and (2) centroid-based saliency exploitation techniques are presented to compute the relative motion vector to the salient region of interest: (1) Extract location of maximum salient points:
\begin{equation}
 \argmax_{x, y}(S(x, y, t)), ..., \argmax_{x, y}(S(x, y, t-W_c))   
\end{equation}
for all saliency maps in $W_c$. \\ (2) Determine the map's centroid $\left(\bar{x}, \bar{y}\right)$ such that:
\begin{equation}
\bar{x} = \frac{m_{10}}{m_{00}}, \bar{y} = \frac{m_{01}}{m_{00}}
\end{equation}
by using the spatial image moments: 
\begin{equation}
m_{pq} = {\sum_x} {\sum_y} x^p y^q S(x,y,t).
\end{equation}
Fig.~\ref{fig:saliency-moments} illustrates saliency maps with varying amount of saliency-emphasized regions and the locations that are picked with respect to the exploitation policy. Note that saliency maps computed from equirectangular projected (ERP) images only contain information about the pan and tilt rotations. These locations are then normalized to [-1,1], converted into the differential domain, and then fed into the S-network.
\subsection{Motion Network (M-network)}
Not only the spatial but also the temporal component can play a significant role for gaze prediction. Moving objects within a scene might temporarily influence the human's gaze direction. Motion maps $M(x, y, t)$ are created with optical flow to quantify the amount of scene motion present within a viewport. Farneback's motion estimation algorithm \cite{Far03} is selected to calculate the actual motion between two consecutive frames. To extract the relative motion within the scene, the ego-motion of the user is removed by subtracting it from the corresponding motion vectors. The resulting dense vector field is then converted into a grayscale motion map that resembles the probability distribution of saliency maps. A Gaussian filter is applied to smooth the map and remove noise. The max-based exploitation policy is leveraged to compute the inputs of the M-network.
\subsection{Deep Spatio-temporal Fusion Network (F-network)}
\begin{figure*}[th]
	\centering
	\includegraphics[width=0.9\textwidth]{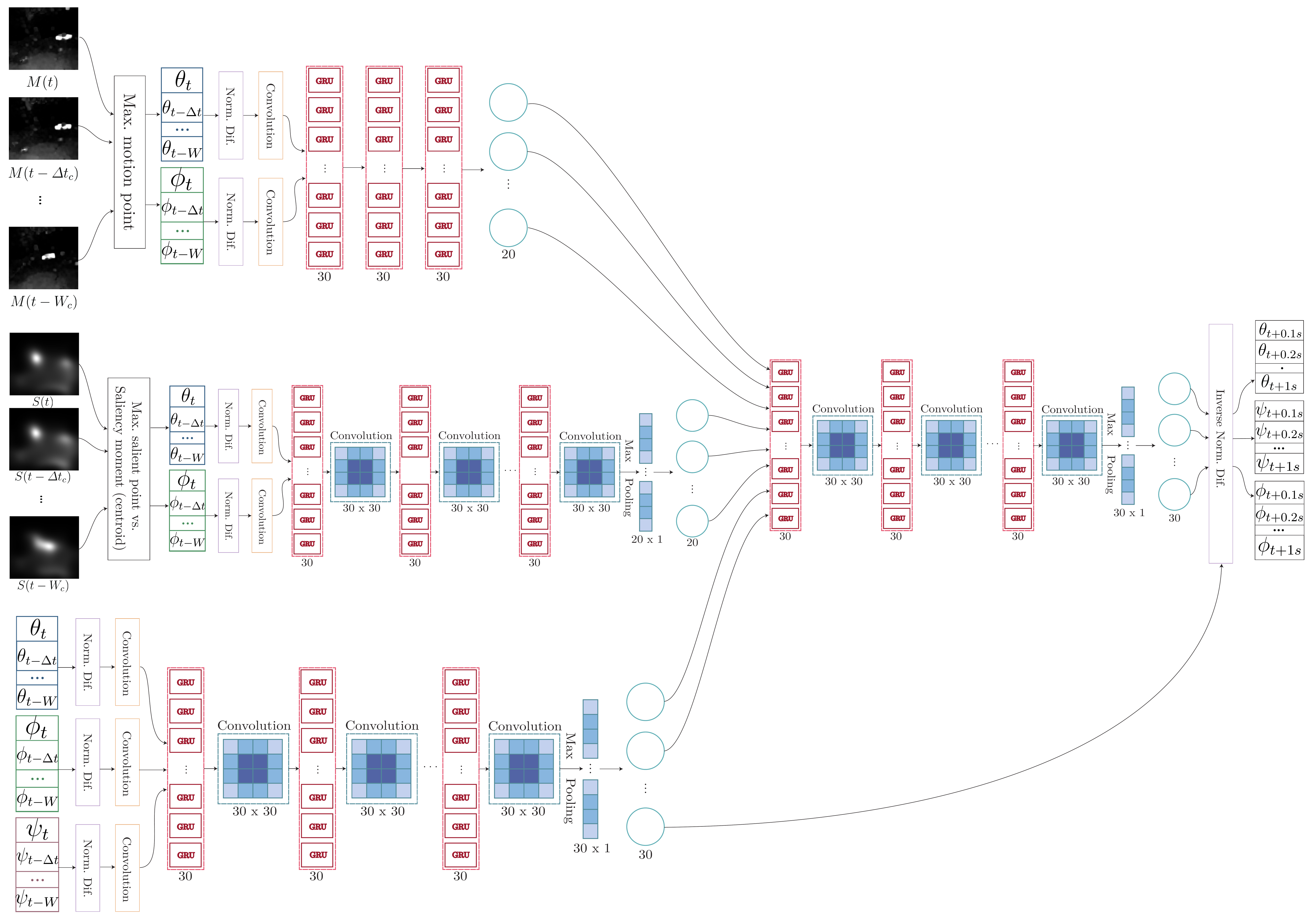}
	\caption{Schematic illustration of the proposed deep late-fusion network (F-network).}
	\label{fig:finalNetwork}
\end{figure*}

The H-, S-, and M-networks are individually pre-trained prior to fusion. Their last layers are then omitted to feed their high level features into the proposed deep fusion network (\textit{F-network}) presented in Fig.~\ref{fig:finalNetwork}.
The F-network consists of an interleaved structure of GRU layers, each with 60 cells, and convolution units with a kernel size of 60$\times$60. A max pooling layer is followed to reduce the dimensions to the desired output size, while preserving important features. A dense feedfoward neural network (FFN) layer with 30 nodes is added to increase the model's non-linearity before remapping them to absolute orientation values by computing their inverse of normalized differences as stated in Equation~(\ref{eq:remapping}).
\subsection{Experimental Setup}
Reproducable and meaningful results are computed by using the IMT dataset with real HM profiles of 58 participants who watched six 360$\degreee$ videos for approximately \SI{70}{\second} \cite{Co17}.  These HM profiles along with their associated image frames are employed to train the proposed networks. 6-fold cross validation is applied to train the network. \SI{60}{\percent} of the videos/users are used for training, \SI{20}{\percent} for validation and testing, respectively. The L1-norm is chosen as loss function. The Adam optimizer is used to update network weights. The batch size is specified to $2^{11}$ with a  learning rate of $10^{-3}$. A learning rate decay scheduler is incorporated to decrease the initial learning rate every ten epochs by \SI{10}{\percent} for a maximum number of 1000 epochs. The early stopping technique is embedded to monitor the validation loss and stop the learning process ahead of schedule with patience of ten epochs to avoid overfitting. During the training process, the core of the individual H-, S-, and M-network is defined to be non-trainable to leave the previously pre-trained weights unchanged.  ReLUs are utilized as activation functions for the dense FFNs. All deep learning architectures and models are designed and trained with the same hyperparameters in Keras and TensorFlow. The fusion network is able to infer future HMs in the domain of microseconds (Core i7 (x64), GeForce GTX 1080 Ti) making it suitable for realtime applications.
\subsection{Discussion}
\begin{figure}[t]
	\centering
	\begin{tabular}{cc}
	\includegraphics[width=.4\textwidth]{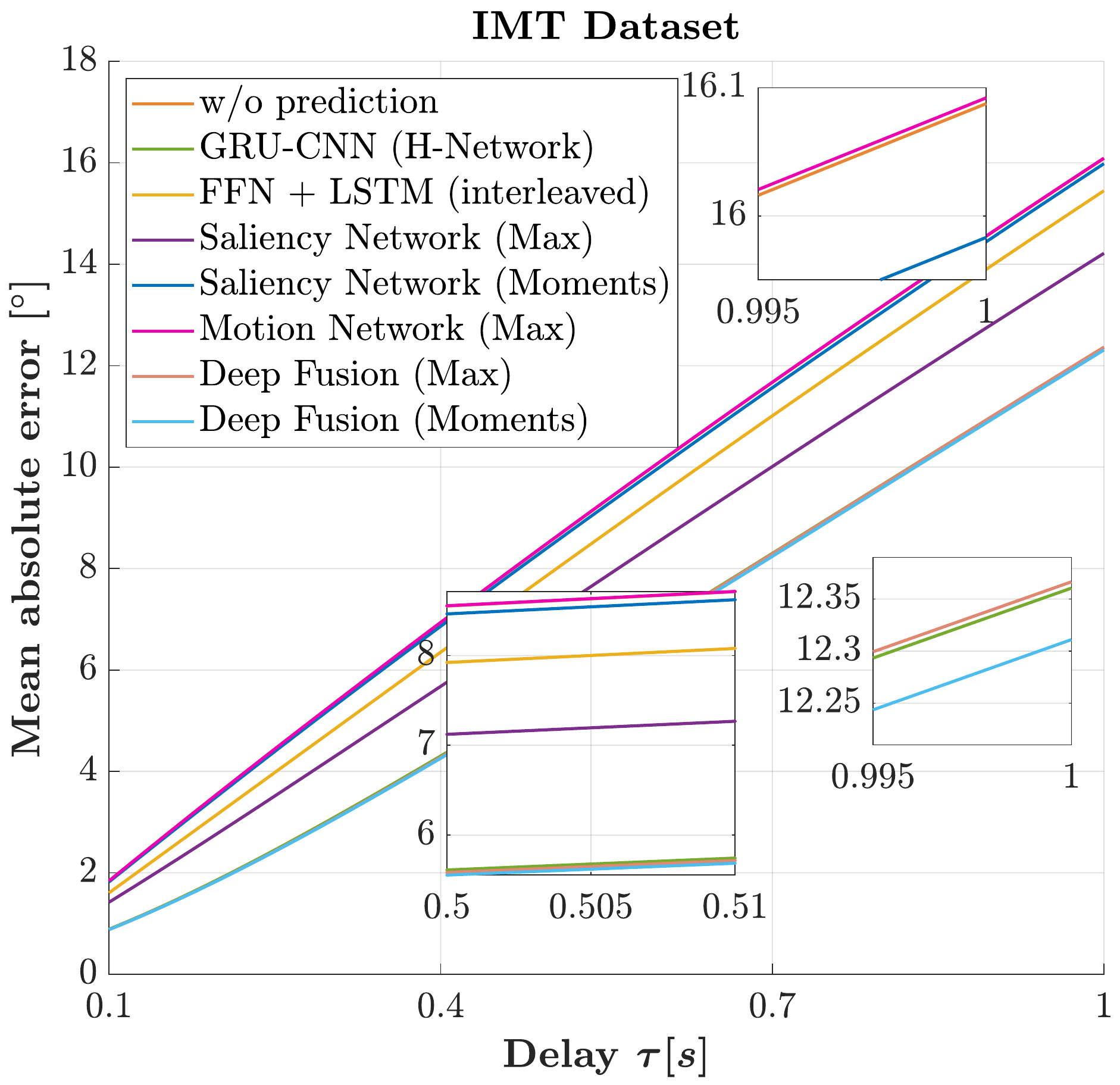} &
	\includegraphics[width=.4\textwidth]{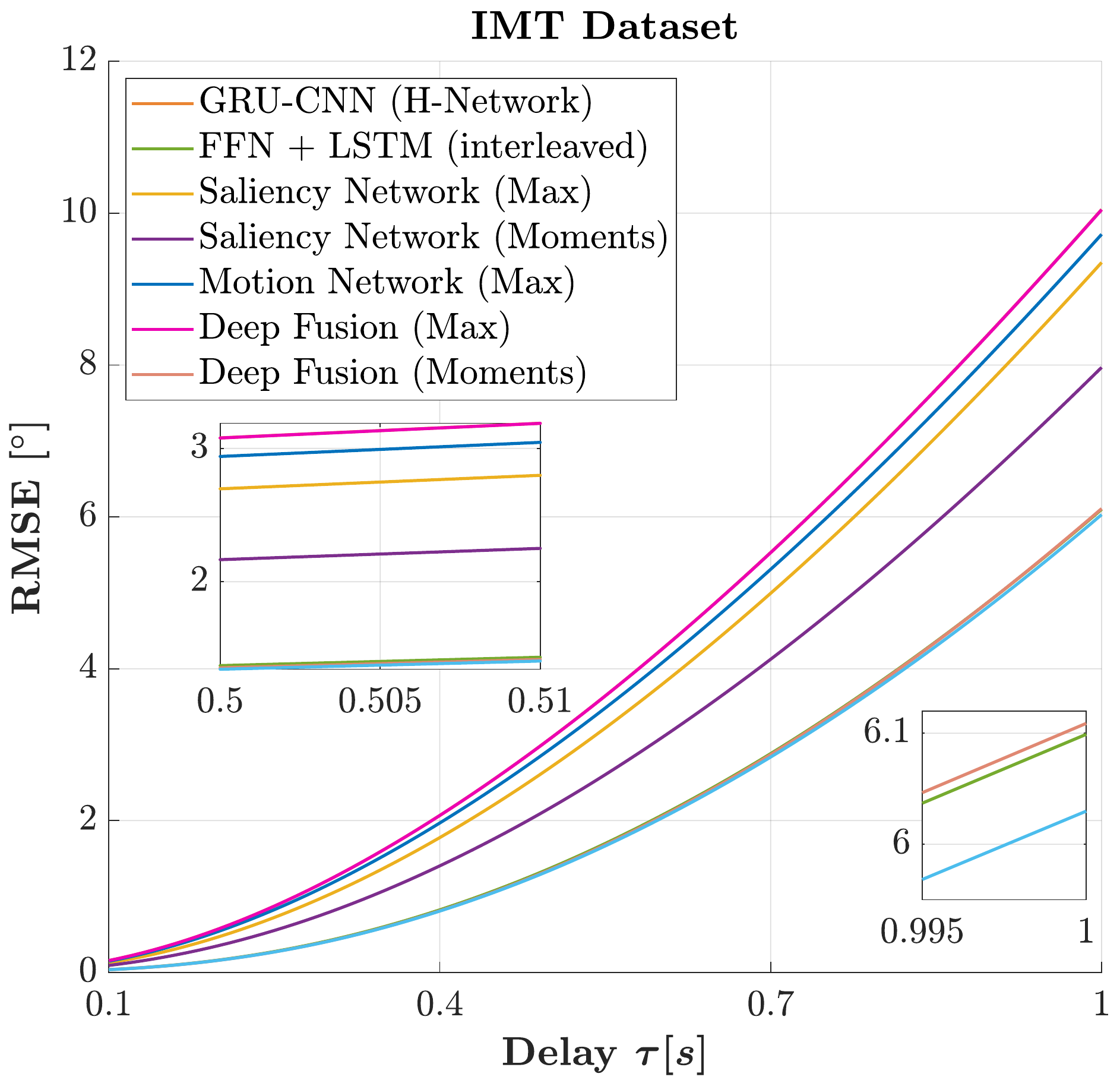}\\
	(a) & (b) \\
	\includegraphics[width=.4\textwidth]{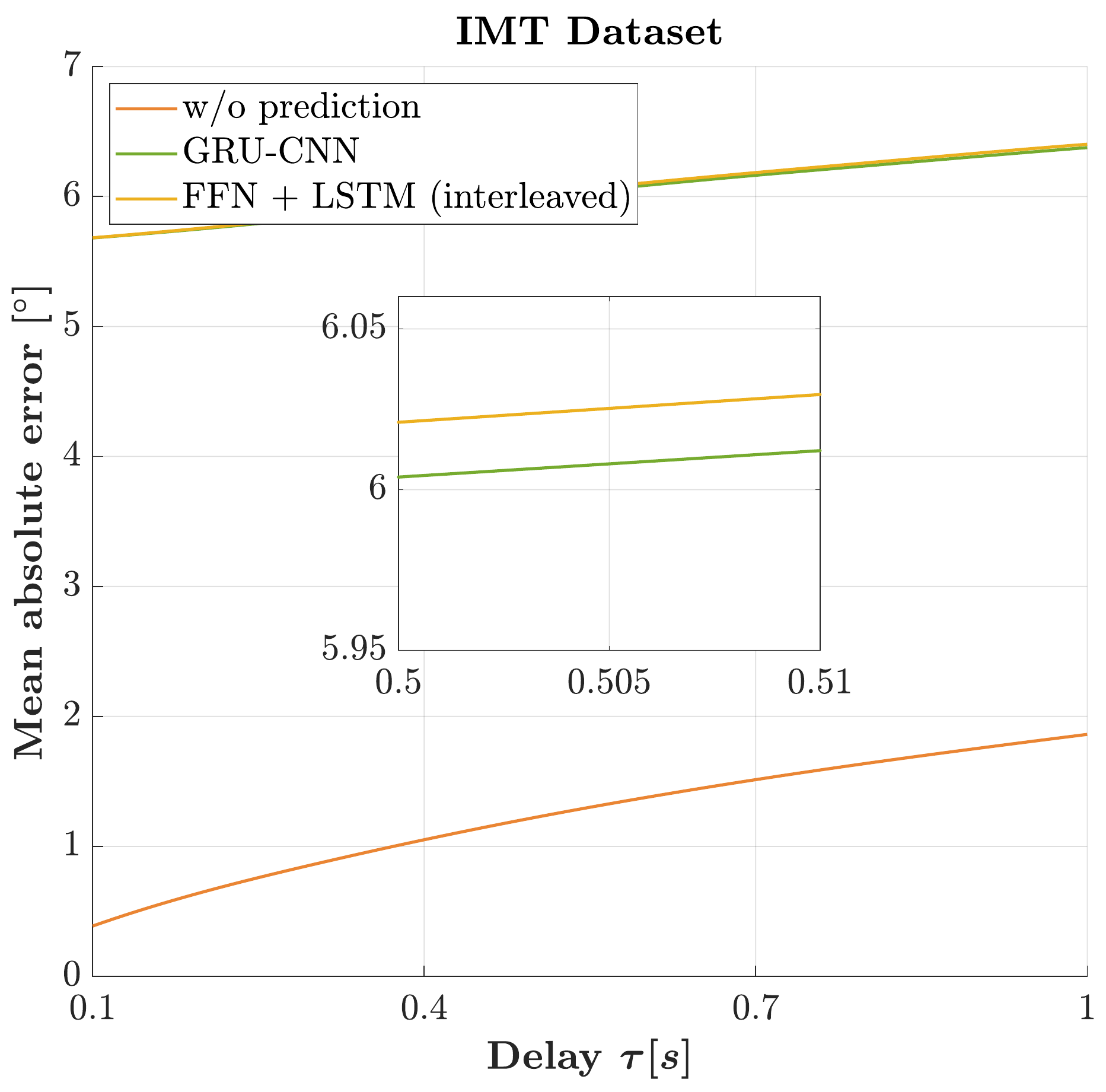} &	
	\includegraphics[width=.4\textwidth]{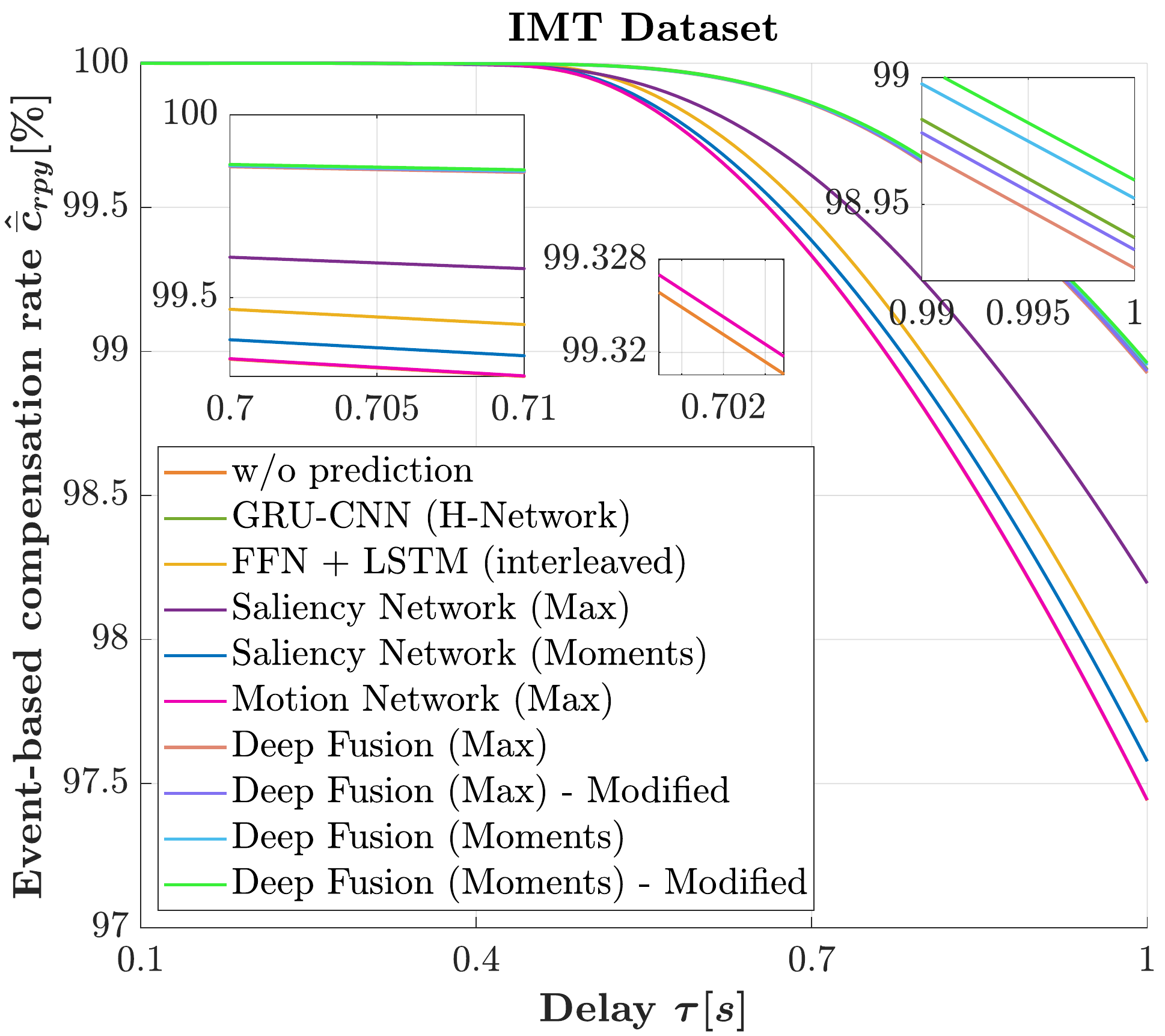}\\
	(c) & (d)  \\
	\end{tabular}
	\caption{The results in (a), (b) convey the accuracy of predicting horizontal HMs and (c) for roll rotations. (d) shows the compensation rates for the investigated prediction models.}
	\label{fig:results-fusion-network}
\end{figure}

Ablation studies are conducted to investigate the isolated contributors of the late fusion model. Their prediction accuracies for horizontal (pan) HMs, which are known to be the dominant ones, are depicted in Fig.~\ref{fig:results-fusion-network}~(a) and (b). The LSTM-based deep network is chosen as representative SotA as it proved superior performance compared to previous work~\cite{TA18b}. 

Merely deploying motion maps turns out to be less rewarding. Its MAE and RMSE values resemble the prediction-less computing. Spatial scene information, instead, appear to be more useful for viewport prediction than the temporal one at pixel level. The isolated contribution of the S-network is inspected for both the max- and centroid-based exploitation policy. The max-based computation achieves thereby considerable accuracies, clearly outperforming the LSTM-based related work. The centroid-based exploitation, however, yields only marginal improvements at the first glance. Computing the centroid is a fairly passive way of exploiting saliency maps, particularly in the presence of multiple saliency-emphasized regions. The max-based approach is more risky and votes for one specific region of interest. Wrong votes, however, lead to diverging estimates between the S- and the H-network resulting in less accurate predictions within the fusion framework as is observable in Fig.~\ref{fig:results-fusion-network}. Best results are achieved when fusing HM-data with moment-based saliency maps. The passive nature of centroids functions thereby as a balance term and prevents the F-network from overshooting. The key driver of the deep fusion network is the H-network, which yields remarkable results independent of scene semantics.

The prediction for roll rotations can only be performed by HM-based deep networks, as saliency and motion maps do not contain any awareness for them. Their accuracies are shown in Fig.~\ref{fig:results-fusion-network} (c)). Surprisingly, both networks clearly fail in predicting roll rotations. Not predicting at all demonstrates even better performance.
This finding is incorporated into the resulting deep fusion framework. The roll rotations are no longer considered for prediction. Instead, the current roll orientation is leveraged as look-ahead value. The M-, S-, and F-networks are modified accordingly.  
The achievable delay-compensation rates for the adopted telepresence scenario \cite{TA19} of all investigated models are shown in Fig.~\ref{fig:results-fusion-network}~(d). The results confirm the superior performance of the centroid-based F-network. Although the max-based S-network is predominant as standalone model in contrast to its centroid-based pendant, the F-network performs better with the moment-based exploitation technique. Considering the current roll rotations as look-ahead values further improved the inference accuracy of the modified F-network. The modified, centroid-based F-network clearly outperforms the SotA and demonstrates a substantial level of delay-compensation. Fig.~\ref{fig:results-final}~(a) compares the best performing prediction scheme to the naive approach, where HM prediction and delay compensation are not deployed. A mean compensation rate of \SI{99.99}{\percent} is achieved for the proposed approach. Even for delay as high as \SI{1}{\second}, a compensation rate of $~$\SI{99}{\percent} is obtained. 
%------------------------------------------------------------------------
\section{Application to On-Demand 360-degree Video Streaming}
The proposed semantic viewport prediction algorithm is not limited to telepresence applications. A proof-of-concept scenario is implemented to highlight the benefits of the proposed approach for on-demand streaming of high-resolution \omni videos, which are characterized by high bandwidth and low latency requirements.
\subsection{Proposed Streaming Framework}
A Dynamic Adaptive Streaming over HTTP (DASH) based communication framework is established to stream on-demand \omni video content between a server and a client as depicted in Fig.~\ref{fig:telepresence-streaming}.  Streaming the complete \omni video footage claims large portions of the transmission capacity although only the current viewport is consumed by the user. A tile-based, viewport adaptive streaming policy is presented to reduce the massive rate requirements and ensure a high QoE by means of proper viewport prediction.
The \omni video is first projected into the 2D ERP format, spatially divided into tiles, and then temporally bundled into time segments of \SI{1}{\second}. Practical experiments proved a $N=4\times8=32$ tiling scheme to be a good trade-off in terms of encoding efficiency and high granularity for superior viewport adaptability. Each chunk is independently encoded (HEVC, main profile) at different bitrates, by varying the quantization parameter (QP) from 22 to 42 with a step size of 5. The server has five representations that can be requested from the client as a function of the available transmission resources. The DASH client is responsible for the adaptation behavior and requests a suitable representation for each tile at the beginning of the segment download. Prior to the segment download, the client determines the target bitrate based on the estimated throughput of the communication network and the video's buffer level \cite{Tia12}. The decision on each tile's bitrate is premised on the proposed viewport prediction algorithm. Higher bitrates are assigned to tiles that lie within the predicted viewport. With respect to the target bitrate and the predicted viewport location, a rate-distortion optimized tile quality selection is performed. The objective is to find the optimal bitrate for each tile to maximize the user's QoE.

The QoE is defined as a measure that describes the weighted distortion and the spatial quality variance, which needs to be minimized for the best user experience. The implemented optimization framework subdivides the image into a grid of $N=32$ tiles. Each tile $n$ has $B$ possible bitrates $b$. $d_{nb}$ and $R_{nb}$ term the level of the present distortion and bitrate of tile $n$ at the $b$th representation, respectively. Tiles that lie within the field of view of the predicted viewport trajectory for the next time segment receive higher bitrates than tiles in the peripheral. A normalized weight:
\begin{equation}
\tilde{w}_n = \frac{w_n}{\sum_{i=1}^B w_i} \hspace{.5em} \text{with} \hspace{.5em} w_n = \frac{1}{\Delta_n^2+1}
\end{equation}
is assigned to each tile to quantify its relevance for bitrate allocation, where $\Delta_n$ denotes the Euclidean distance of a tile $n$ to the center of the closest tile that belongs to the prospective viewport. Given the ERP image in the $YUV$ format, the distortion $d_{nb}$ is computed as the mean square error (MSE) between the $Y$-frames of the original and the reconstructed tiled videos $\hat{Y}$:
\begin{equation}
d_{nb} = \frac{1}{\text{w}\cdot\text{h}}{\sum}_{i=1}^{\text{w}} {\sum}_{j=1}^{\text{h}}(Y(i,j)-\hat{Y}(i,j))^2,
\end{equation}
with w and h denoting the tiles' width and height. Note that the distortion within an ERP tile does not accurately reflect the perceived distortion by the user. The actual distortion:
\begin{equation}
    d'_{nb} = d_{nb}\cdot c_n  
\end{equation}
of each tile  with respect to the viewing sphere is instead computed by weighing the distortion $d_{nb}$ with a value $c_n$
to account for the projection effects:
\begin{equation}
c_n = \left(\frac{\pi R}{4}\right)^{-2} \cdot {\int}_{\phi_n}^{\phi_n+\frac{\pi}{4}}{\int}_{\theta_n}^{\theta_n+\frac{\pi}{4}}R^2\cos\left(\phi\right)d\theta d\phi,
\end{equation}
with $R$ being the sphere's radius. To ensure a satisfying QoE, the \textit{weighted distortion} $\Lambda$ and the \textit{spatial quality variance} $\Xi$ are introduced as measures to indicate the perceivable spatial smoothness:
\begin{align}
\Lambda~&=~{\sum_{i=1}^{N}}{\sum_{j=1}^{B}} \tilde{w}_i\cdot d'_{ij}~\cdot~ a_{ij},\\
\Xi&={\sum_{i=1}^{N}}{\sum_{j=1}^{B}} \tilde{w}_i\cdot |d'_{ij}-\Lambda|\cdot a_{ij},
\end{align}
where $a_{ij}=1$ if tile $i$ is selected to be streamed at the $j$th quality. The overall function that is to be optimized is defined as their conical sum (with weight $\nu$):
\begin{gather}
\min \hspace{.5em}\Lambda + \nu \cdot \Xi \\
\text{s.t.} \hspace{.5em}{\sum_{i=1}^{N}}{\sum_{j=1}^{B}} \mathcal{R}_{ij} \cdot a_{ij} \leq \mathcal{R}_\text{target} \hspace{.5em} \\ 
\text{and} \hspace{.5em} {\sum_{j=1}^{B}} a_{ij}=1 \hspace{.5em} \text{with} \hspace{.5em} a_{ij} \in \{0,1\}.
\end{gather}
The two constraints ensure the overall sum of bitrates to be less than or equal to the target bitrate $\mathcal{R}_\text{target}$ and that only one representation is selected per tile.

To apply SotA optimization solvers, the underlying optimization problem, which uses a nonlinear function for the spatial quality variance, is converted to an integer linear programming (ILP) problem with quadratic constraints. First, an auxiliary variable:
\begin{equation}
 k_{ij}=d'_{ij}-\Lambda   
\end{equation}
is introduced to reformulate the spatial quality variance to:
\begin{equation}
\Xi = {\sum_{i=1}^{N}}{\sum_{j=1}^{B}} \tilde{w}_i\cdot k_{ij}\cdot a_{ij}.
\end{equation}
Further constraints are set to ensure the absolute value of $k_{ij}$: 
\begin{equation}
k_{ij} \geq d'_{ij}-\Lambda \hspace{.5em} \text{and} \hspace{.5em} k_{ij} \leq -\left(d'_{ij}-\Lambda\right).
\end{equation}
Another auxiliary measure:
\begin{equation}
    l_{ij}= k_{ij}\cdot a_{ij}
\end{equation}
is used to remove the nonlinear multiplication: 
\begin{equation}
\Xi = {\sum_{i=1}^{N}}{\sum_{j=1}^{B}} \tilde{w}_i\cdot l_{ij}
\end{equation}
obtaining an optimization problem with quadratic constraints.
\begin{figure}[t]
	\centering
	\begin{tabular}{cc}	
	\includegraphics[width=.4\textwidth]{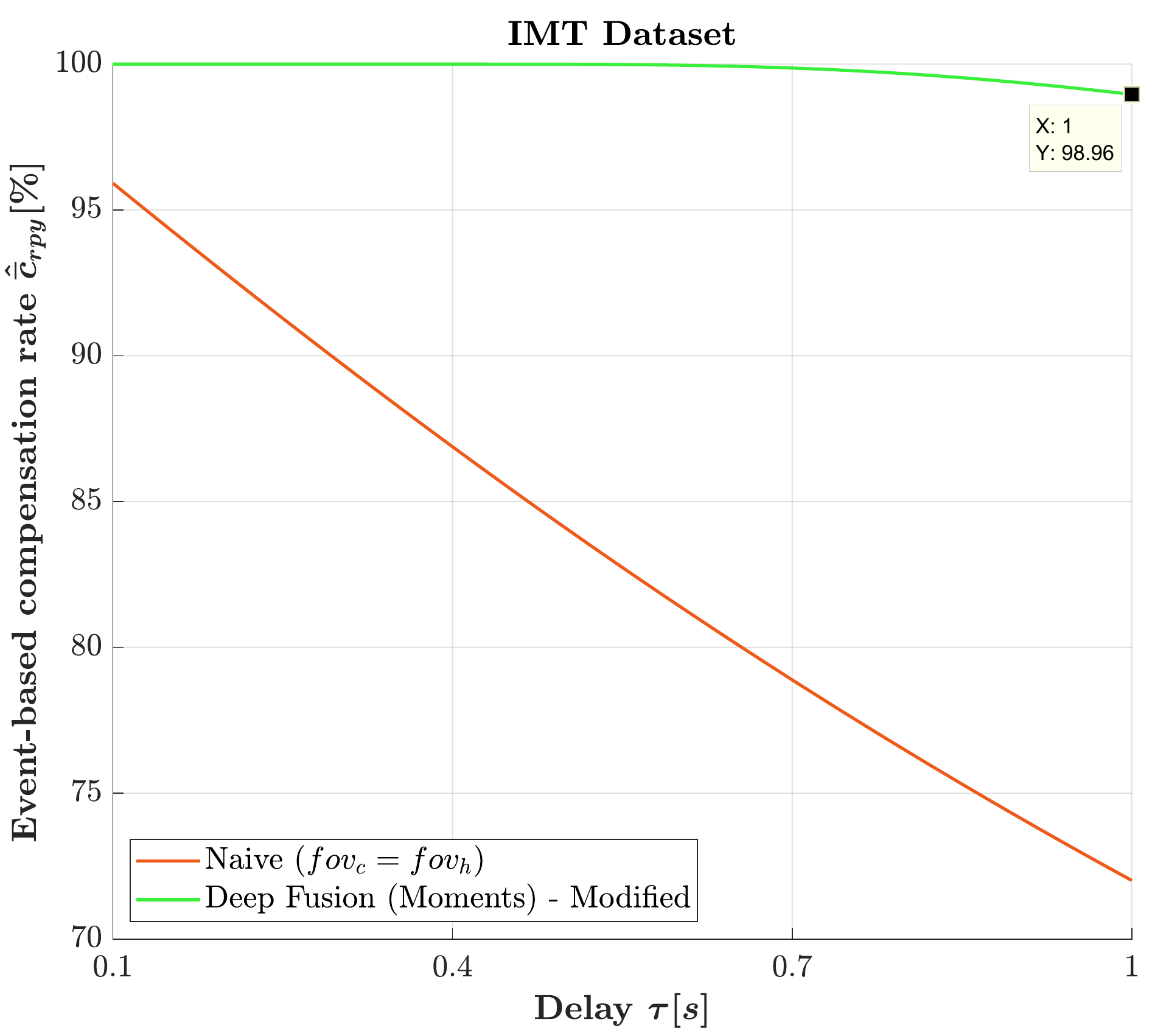} &
	\includegraphics[width=.4\textwidth]{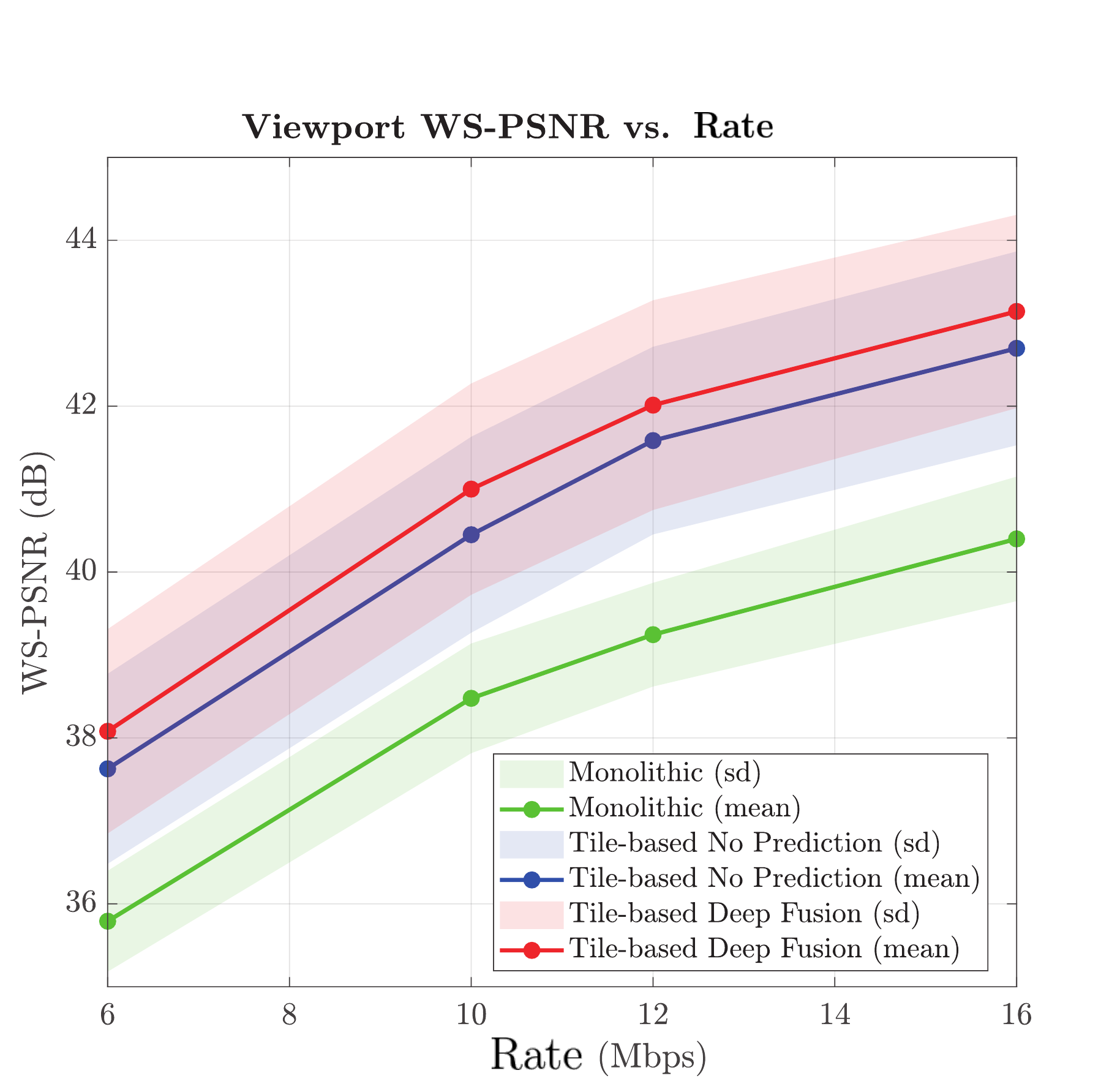}\\
	(a) & (b) 
	\end{tabular}
	\caption{(a) Results of semantic viewport prediction vs. the naive approach with no prediction. (f) The perceived video quality in terms of WS-PSNR over the consumed transmission rate.}
	\label{fig:results-final}
\end{figure}
The impact of the proposed semantic viewport prediction scheme on the perceived video quality is quantified with the peak signal-to-noise ratio (PSNR). Computing the PSNR within the ERP does not directly reflect the perceived video quality due to projection artifacts. Rather than computing the reprojected viewport for every HM and subsequently calculating the PSNR, a computationally more efficient method is given by the weighted-to-spherically-uniform (WS-)PSNR \cite{Sun17}. The WS-PSNR accounts for projection artifacts and computes the viewport PSNR by using the viewport pixels within the ERP plane. This is done by assigning projection-dependent weights to each pixel $(x,y)$~\cite{Sun17}:
\begin{equation}
w_\text{ERP}(x,y) = \cos\left(\left(y-\frac{h_\text{ERP}}{2}+\frac{1}{2}\right)\cdot\frac{\pi}{2}\right),
\end{equation}
where $h_\text{ERP}$ denotes the ERP plane's height. Due to the oversampling effect of the ERP towards the pole regions, weights are decreased from the equator to the poles of the ERP plane.
The WS-PSNR for an image $I_\text{ERP}$ in the ERP format can then be computed as:
\begin{equation}
\text{WS-PSNR} = 10\log\left(\frac{\left(\max(I_\text{ERP}(x,y)\right)^2)}{\text{WMSE}}\right),
\end{equation}
by utilizing the weighted MSE (WMSE) for a viewport $V$ as:
\begin{equation}
\text{WMSE}= {\sum_{(x,y)\in V}} \frac{(Y(x,y)-\hat{Y}(x,y))^2 \cdot w_\text{ERP}(x,y)}{\sum\limits_{(x,y)\in V} w_\text{ERP}(x,y)}.
\end{equation}
\subsection{Discussion}
Subject studies in prior art proved already that a delay compensation approach results in a more immersive and pleasant user experience. In this paper, we inspect the WS-PSNR with respect to the consumed transmission rate as an objective qualitative measure. The average viewport WS-PSNRs as a function of the transmission rates are shown in Fig.~\ref{fig:results-final}~(b). The mean and standard deviations are visualized by means of the deployed 6-fold cross-validation. The network throughput is kept constant during streaming. Multiple sessions are carried out for different throughput values, which range from 6--16 Mbps. The WS-PSNR is thereby examined for the monolithic streaming, where the whole \omni video footage is sent, compared to the tile-based approach, and its extension with the proposed viewport prediction paradigm. The results confirm that tile-based streaming performs better than monolithic streaming. Adopting the proposed viewport prediction technique clearly demonstrates further improvements. Significantly higher viewport WS-PSNR values are achieved when deploying tile-based streaming along with proper viewport prediction.
%------------------------------------------------------------------------
\section{Conclusion}
Viewport prediction is widely deployed to improve the QoE of immersive streaming applications and to lessen the stiff burdens on the communication network. A late-fusion scheme is presented that merges head motion data with spatio-temporal scene content, outperforming related work both for immersive telepresence and on-demand \omni video streaming.

Providing a single prediction model that is supposed to work for all individuals in the world despite the varying anatomy and motion habits of each person is challenging. In future work, we plan to provide an algorithm that is able to adapt to each person independent of the task at hand. %This can be achieved by devising an offline/online prediction policy that uses a pre-trained base model as foundation and customizes the network’s weights through an online learning process. 

\section{Acknowledgement}
This work has been supported by the Max Planck Center for Visual Computing and Communication.

\bibliographystyle{unsrt}  
\bibliography{ms}  %%% Remove comment to use the external .bib file (using bibtex).

\end{document}